\documentclass[useAMS,usenatbib]{mn2e}
\usepackage{graphicx}
\usepackage{txfonts} 

\newcommand{\aap}{A\&A}

\newcommand{\mnras}{MNRAS}
\newcommand{\apj}{ApJ}

\newcommand{\apjs}{ApJS}

\catcode`\@=11
\def\gsim{\ifmmode{\mathrel{\mathpalette\@versim>}}
    \else{$\mathrel{\mathpalette\@versim>}$}\fi}
\def\lsim{\ifmmode{\mathrel{\mathpalette\@versim<}}
    \else{$\mathrel{\mathpalette\@versim<}$}\fi}
\def\@versim#1#2{\lower 2.9truept \vbox{\baselineskip 0pt \lineskip
    0.5truept \ialign{$\m@th#1\hfil##\hfil$\crcr#2\crcr\sim\crcr}}}
\catcode`\@=12
\def\msun{\hbox{$M_\odot$}}

\def\yr-1{\hbox{${\rm yr}^{-1}$}}

\def\t9{\hbox{$t_9$}}

\def\m*{\hbox{$M_{\rm stars}$}}

\def\ho{\hbox{$H_\circ$}}
\def\h50{\hbox{$\ho /50$}}

\begin{document}

\title{ Why stars inflate to and deflate from red giant dimensions, II: \\ replies to critics}
\author[Alvio Renzini]{Alvio Renzini$^{1}$\thanks{E-mail: alvio.renzini@inaf.it}\\ 
$^{1}$INAF - Osservatorio
Astronomico di Padova, Vicolo dell'Osservatorio 5, I-35122 Padova,
Italy}

\date{Accepted 2023 January 10. Received 2022 December 16}
\pagerange{\pageref{firstpage}--\pageref{lastpage}} \pubyear{2002}

\maketitle
                                                            
\label{firstpage}

\begin{abstract}
In a 1992 paper of ours the role of opacity-driven thermal instabilities in shaping the course of stellar evolution was amply illustrated. This included the classical issue of
``{\it why stars become red giants"} as well as the subsequent formation of extended  ``Cepheids" {\it blue loops} during the helium burning phases. Our explanation of these evolutionary phenomena has been occasionally dismissed with just a few words in refereed or not refereed publications. In a most recent case, the fact that, through the years, I did not reply to these criticisms  is interpreted as evidence that they were well founded. In this paper it is made clear that this is not at all the case,  the leading role of such instabilities is instead reaffirmed and the criticisms are shown to be insubstantial. 

\end{abstract}


\maketitle

\section{Introduction}
\label{sec:intro}
The founding fathers of stellar evolution, Subrahmanyan Chandrasekhar, Martin Schwarzschild, Alan Sandage and Fred Hoyle, set up the stage, made the early discoveries and introduced the {\it language} of stellar evolution. But left a problem unsolved, which is usually formulated as a question: ``why stars {\it become} red giants?"$^2$\footnotetext[2]{The reason to use italics fonts for this verb will be made clear in the course of the paper.} Then, in the early 'sixties, adequate computers became available and, thanks to scientists such as Rudolf Kippenhahn, Pierre Demarque and Icko Iben, full realistic stellar evolutionary sequences began to be produced. Computers had no difficulty whatsoever to make red giants, and to account, for instance, for the so-called Hertzsprung gap, i.e., the paucity of stars in between the main sequence (MS), spectral types O to F, and the red giants branch (RGB), spectral types K and M. Models swept quickly through to the gap and found a rest only as red giants. Problem solved?
Yes and no. The clear success of the models satisfied almost everybody, and the question {\it Why?} became kind of academic and few continued to pay attention to it. A common wisdom was that
it is not easy to predict, without computers, what  the solutions are of a system of four, non-linear, differential equations.

A first hint came to me when realising what was happening in the evolution of a $3\,\msun$ stars in one of the classical Iben's papers \citep{iben65}.  In its Figure 6, one can see that some 10 Myr after hydrogen exhaustion in the core, the surface luminosity of the star was starting to drop while the nuclear luminosity (provided by the hydrogen burning shell) was still increasing for another $\sim 2$ full Myr, and only then it started to drop as well. So, something was happening to the envelope that the core did not know yet. The envelope had become incapable to transmit outwards and radiate away all the energy being generated in the interior, i.e., the envelope had become thermally unstable!
This incapacity was clearly causing radiative energy being trapped in the envelope, which in turn was causing its runaway expansion.
Thus, a thermal instability of the stellar envelopes was first proposed as the driving of the stars to red giant dimensions \citep{renzini84, iben84} This was the beginning of the story.

\section{Runaway inflations and deflations of stellar envelopes}
\label{sec:why}

This explanation  was more effectively formulated in a paper with the same title of the present one \citep{renzini92}, along with the presentation of stellar evolutionary sequences and toy envelope models illustrating what physically happens to the stars.$^3$\footnotetext[3]{Most of these calculations were made with Iben's stellar evolution code that Icko 
kindly provided to us.} I could find no better way for summarising the paper than use its very abstract, which is reproduced here:

``We demonstrate that a unique physical process is responsible for the runaway expansion of stars to red giant dimension, as well as for their subsequent recollapse leading to the so-called blue loops. In response from an increasing luminosity from the core, the stellar envelope expands keeping its thermal equilibrium$^4$\footnotetext[4]{A star is said to be in thermal equilibrium when its surface luminosity remains close to the nuclear energy generated in the interior.} insofar the envelope thermal conductivity increases. However, expansion implies local cooling, ion recombination, and thus increasing opacity, in such a way that a time comes when further expansion causes a drop in thermal conductivity in the envelope. As the luminosity transferred outward and radiated away from the surface drops, thermal equilibrium is broken and an increasing fraction of the core luminosity is trapped in the envelope, causing further expansion and further drop of the thermal conductivity: the resulting runaway inflation of the envelope brings the star into the red giant region of the H-R diagram, and thermal equilibrium is not restored until convection penetrates inwards and the whole envelope becomes convective. The reverse process is responsible for the formation of the blue loops. During the early helium-burning phase, the core luminosity decreases and the star descends along the Hayashi track. By contraction the envelope heats up, the heavy elements ionize and the opacity drops. As the inner part of the envelope returns to radiative equilibrium, the envelope departs again from thermal equilibrium, since by contraction the temperature increases, the heavy ions ionize, the opacity drops, the thermal conductivity increases, and so do the radiative energy losses. Thus the envelope catastrophically deflates inside the potential well of the star. We present detailed analyses of these runaway inflations and deflations, and apply those concepts to achieve a deeper physical understanding of several major features of stellar evolution, etc...''

So, from the beginning ours was not just an explanation of why stars {\it become} giants, but also of why some stars retreat from a red giant size and experience extended blue loops during their core-helium burning phase, another unexplained issue at that time. \cite{lauterborn71} had indeed realized that what they called a ``secular instability" was developing ``during" the loops, but could not identify its physical origin. Our 1992 paper has been often criticized, but only for the red giant issue, while the loops part was ignored. This was somehow disappointing, because ours ambition had been to {\it unify} the two problems as a manifestation of a single process: a thermal instability arising in stellar envelopes, where opacity plays a leading role. For many years I left unanswered the criticisms, hoping that time would have settled the issue, but this did not happen. Once every few years a paper appears cursorily dismissing our explanation and venturing into attempts to find the real one, but never claiming success. The latest case is a paper by \cite{miller22} that has recently appeared on the ArXiv, where one reads: ``The discussions about why stars become red giants have sometimes turned into heated debates (\citealt{sugi97, faulkner97, sugi00, faulkner05}), while some other times authors have ignored criticism, and continue to develop ideas (\citealt{renzini92, renzini94}) that had already been seriously questioned by other researchers (\citealt{weiss89, iben93})". So, our explanation was dismissed on the basis that others had already claimed  that it is bogus and the fact that we did not further react is interpreted as an implicit admission that the criticism was well founded. So, this paper is meant to make clear that this is not at all the case. I still believe that the explanations in  our 1992 paper are correct and describe why (some) stars {\it become} red giants, and others don't.


\section{The critics}
\label{sec:critics}
In this section the criticisms made by the mentioned authors are quoted and comment. Let's start from \cite{weiss89}. In his paper Achim Weiss examines the criterion for thermal stability of stellar envelopes that I had proposed in 1984 and he found some cases in which apparently it would not apply. In particular, an approximate rendition for the criterion was found inaccurate. This last point was admitted in our 1992 paper and the other criticisms where answered. But in the end Weiss concluded: ``I agree with Renzini's discussion on the differences in redward evolution (Renzini 1984) of stars of different masses, different metal content and his conclusion about the connection with stellar opacities. Also, this paper confirms that the expansion is a pure envelope phenomenon that has to be initiated somehow (e.g. by a rapid contraction of the core and the presence of a burning shell). It might be acceptable that the final expansion in some stars after having reached a maximum luminosity is indeed a thermal runaway effect connected with some specific opacity features." So, rather than demolish it, Weiss confirmed the proposed physical origin of the inflation to red giants.

Sugimoto \& Fujimoto (2000) confine their criticism to the following sentences: ``Renzini et al. (1992) and Renzini \& Ritossa (1994) have argued that thermal instability in the envelope, which results in the deep penetration of surface convection, is the cause of the expansion to a red giant. It is true that the red giant branch is separated from the branch of stable blue giants without surface convection by the well-known Hertzsprung gap, as first noted by \cite{kozlowski71} and \cite{lauterborn72}  and also seen from Figure 7. A star does undergo thermal instability in the envelope when crossing the gap. This is not the cause of the large expansion, however, but is a result of potentially extraordinarily large radii, as already noted by many authors (\citealt{yahil85, whitworth89, fuji91, iben93})". 
So, Sugimoto and Fujimoto concur that a thermal instability develops in the envelope, however it would not be the thermal instability that drives the expansion but rather the instability would be driven by the  {\it willingness} of the star to assume a stable giant configuration that terrestrial mathematicians had proved to exist. I don't know how to otherwise interpret this sentence.

According to Fujimoto \& Iben (1991):  ``Renzini (1984) discusses the thermal instability in the envelope as a cause of the red giant structure."  This sentence does not properly
represent what said in my 1984 paper, where the thermal instability is seen the cause of the {\it expansion} to the red giant structure, rather than of the existence of static red giant solution to the four equations. Then Fujimoto and Iben go on: ``... during the evolution to a red giant, a thermal instability in the envelope occurs and the surface convective zone extends deeply into the interior... however, the envelope instability is just an episode which occurs when the surface temperature grows low enough in the course of expansion, forcing the helium ionization zone inward (in the case of helium stars) or forcing the hydrogen and helium ionization zones inward (in the case of hydrogen stars), and causing the effective polytropic index to increase. This instability is not itself responsible for the development of a red giant structure (see also Weiss 1989)." However, this is not what Weiss concluded, as reported above, if by ``development" one means {\it expansion} of a dwarf to become a giant. If instead they meant {\it existence} of a red giant structure solution to the equations, then the same comment to the point of view of Sugimoto \& Fujimoto applies, as above. Incidentally, in our 1992 paper it is shown that  through the expansion the polytropic index in the envelope does not change much at all (see Figure 9 there).

Iben (1993) had four objections to our 1992 paper,  partly following the lines of Fujimoto \& Iben (1991). Contrary to what stated by Miller Bortolami (2022), we did not leave unanswered the four objections, but in \cite{renzini94} we demonstrated all the four points to be invalid, so there is no need to expand here and the interested reader may look directly at Iben's views and our replies.


\section{STATIC STELLAR MODELS}
\label{sec:static}
In the course of their evolution, stars either expand or contract all the time. Why are they doing so? On the MS and beyond, what drives stellar evolution is nuclear burning in the deep interior, whereas the envelope has a passive role; it has {\it just} to adjust to the changing conditions in the core. However, details of the core internal structure are irrelevant, as there is no {\it underground} communication between the envelope and the core. What matters to the envelope are just three core quantities, namely, its mass, radius and luminosity, which set the the boundary conditions at the base of the envelope. Of the three, by far the the dominant one for the driving of envelope changes is luminosity. For example, take a star during its core hydrogen-burning phase: in the conversion of hydrogen into helium three particles disappear, pressure would drop (because $P=nkT$), so the core by loosing pressure support is forced to contract. In doing so it heats up thus being able to maintain its hydrostatic equilibrium. As core density and temperature increase, so does the nuclear luminosity, and this is the trigger to envelope expansion. Suppose the envelope does not change at all while the nuclear luminosity increases. The envelope would receive from its bottom more energy than it is able to emit from the surface, hence its total energy (thermal plus gravitational) would increase: it expands. In general, when the core luminosity increases the star expands, when it decreases the star contracts (at least insofar the envelope is close to thermal equilibrium). Anyway, during these changes deviations from hydrostatic equilibrium are tiny and indeed we speak of {\it quasi-static}  evolving configurations.

Several papers devoted to the issue ``why stars {\it become} red giant?" deal instead with fully static configurations, i.e., having set to zero the gravitational energy generation (the so called $\epsilon_{\rm g}$). This is equivalent to freeze evolution and explore instead potential stellar structures in full thermal equilibrium that may or may not be realized in Nature. As such, these studies may answer the question ``what is the structure of red giants", but having switched off evolution they are by construction incapable of answering the other question, i.e., how such structure {\it becomes} established in real stars, and in stellar evolutionary sequence alike? The key point is that the word {\it become} implies {\it evolution}, not static, unevolving states. The transition to red giant can only be understood in a full evolutionary context.

Most of the papers mentioned in the previous section deal instead with static models, and thus are intrinsically unable to answer the usual question. Fairly enough, \cite{whitworth89} entitled
his paper ``Why red giant are giant". This is indeed a question that static models can answer. But intrinsically cannot account for evolutionary transitions. Not only static models were used, but a specific subset of them (e.g., \citealt{yahil85, fuji91,sugi00,faulkner05}). Perhaps with the ambition to succeed where the founding fathers have failed, they tried to use their same old tools, specifically the (in)famous homology invariants $U$ and $V$. For what matters here, suffice to say that $U$ and $V$ are built using only two of the four equations, namely those for hydrostatic equilibrium and mass conservation. No energy generation. No luminosity, i.e., no energy transfer through the star. Hence, in such approach the drivers of evolution are expunged altogether. I dare to say that virtually nothing about stellar evolution
can be understood by cruising the $UV$ plane. Well, in absence of better tools the founding fathers got something out of it. Not much, but something.
However, after $\sim\! 60$ years of full stellar evolution through computers the regression to the rudimentary tools of the fathers sounds inexplicable to me.
Apparently, there was a kind of reluctance to look at what in fact happens inside stars from computer outputs, as if knowledge achieved in that way was lacking some sort
of sublimity. Perhaps these feelings were best expressed by \cite{faulkner05}, when saying ``The end result is that the post-main-sequence developments of all stars 
-low- mass, intermediate-mass and high-mass- as they expand to become giants, are finally seen to be example of one underpinning fact: that dense cores with surrounding shells naturally follow hydrogen exhaustion. While this has been known all along from oft-repeated computer calculations, we now know why analytically. That matters to true theorists." (p. 150). 


\section{A chicken and egg problem?}
\label{sec:egg}
 \cite{miller22}$^5$\footnotetext[5]{It has been thanks to this paper that I discovered Faulkner's article of 2005, as it never appeared on ArXiv.} fully endorses Faulkner's (2005) criticism of our physical explanation of the 
dwarf-to-giant transition, so I feel obliged to comment on such criticism. I cut and paste here all his points relative to us, and comment them.

``For the most part, such stars [massive post-main-sequence stars] are out of thermal equilibrium, but not thermally unstable as Renzini has claimed, alone or in concert, in a series of papers, e.g. Renzini et al. (1992)."  (p. 190) In his whole article  there is no attempt at elucidating what would be the difference between being out of thermal equilibrium and being thermally unstable, specifically  in stars expanding to red giants. In the early hydrogen shell-burning phase stars can be in or very close to thermal equilibrium for quite some time (see the run of nuclear and surface luminosity in Iben's Figure 6), until suddenly they start to depart precipitously and increasingly from thermal equilibrium. What is this if not a thermal instability? 

A few pages later, he states: ``We now see that it is most definitively not an envelope thermal instability that drives the expansion of the entire star, as Renzini, either alone or with a succession of (sometimes absconding) colleagues has long asserted. To the absolute contrary, the envelopes are sent out of equilibrium by an expansion imperative dictated by developments in the deepest and densest interiors of the stars." (p. 196).

Then Faulkner goes on: ``We can now contrast this with Renzini's proposed explanation for the expansion of stars to the red-giant branch, an explanation he advanced particularly for intermediate mass stars. Renzini suggested that a star of intermediate mass starts to expand from the main sequence - a fact taken as a given - recombination occurring in its cooler regions increasing the opacity there, leading to absorption of luminous energy from below. That in turn expands the star's outermost layers still further, leading to more cooling and yet more absorption, as the hypothesized recombination wave sweeps inward through the star. A thermal instability develops from the outside inwards, for which the luminosity dips are advanced as evidence of this absorption of flux in the outermost regions. 
He has furthermore claimed that this opacity effect is naturally smaller when there are fewer heavy elements in a star, and that this is why such stars do not expand as much.  In this explanation, once the initial expansion from the main sequence has made the outermost layers cool enough, the natural behavior of those layers takes over and promotes the major, fast transition to the red-giant branch itself. Thus the behavior of the outermost layers leads the star to the giant branch.''

This rendition of our interpretation of the phenomenon of the evolution to red giants (for intermediate mass stars) is fair enough, though Faulkner doesn't spare us a couple of pricks$^6$\footnotetext[6]{Does anybody doubt that stars expand during their MS phase? Or, are we the first to {\it claim} that metals contribute to stellar opacity? That ion recombination takes place was not an``hypothesis" but it is a fact that can be easily verified on stellar evolutionary sequences.}. 

But then Faulkner continues:
``I have shown above, instead, that the behavior of the central regions is the main driver. Far from leading the rest of the star along, the envelope regions are {\it lagging behind} where they would have been had there been time for them to reach either a complete new equilibrium mandated by the change in central conditions (in particular the increasing central condensation) or some `moving target' analogue." (p. 199-200)

Here Faulkner is half right, hence half wrong. It is obviously true that the ultimate driver of stellar evolution are the nuclear transformations taking place in the core (including the shell). It is the ensuing increase in luminosity emanating from the core that drives the initial expansion of the envelope, during the first times after hydrogen shell ignition. But once the thermal instability suddenly erupts, it is the envelope that takes the lead. As mentioned in Section I, it is the luminosity radiated by the envelope that starts dropping, while the shell (nuclear) luminosity still keeps increasing for a while. Then, with some delay, also the nuclear luminosity starts dropping fast, which is due to the feedback effect 
from the envelope. As documented  in our 1992 paper, with the runaway expansion of the envelope, its weight on the shell drops, i.e., when the accelerated expansion started from the surface sweeps inside all the way to reach the upper part of the shell, then density and temperature in the upper shell drop and so does the generation of nuclear energy.$^7$\footnotetext[7]{This is what \cite{iben65} called the ``shell narrowing phase", see Iben's Figure 6.}
In this phase the envelope is clearly {\it leading} and the core (the shell) is {\it lagging}, to the extent that the envelope expansion almost succeeds in switching off the shell.

Faulkner's argument seems instead to consist in that the core would be changing too rapidly for the envelope to follow, and then the envelope would lag behind. Hence, there would be no genuine thermal instability in the envelope. But this not what happens, especially in intermediate-mass stars: Past shell ignition there is a long period during which the core shrinks slowly,  nuclear and surface luminosities are almost identical, and yet the thermal instability suddenly erupts (see again Iben's Figure 6). Moreover, in our 1992 paper we already countered this option: in Figure 4 there we showed indeed that the thermal instability erupts, no matter how slowly the core luminosity increases. In other words, it is not that the core would be  evolving too rapidly for the envelope to adjust, it is instead that the envelope becomes incapable of transferring outwards all the energy being provided by the core. However, in stars more massive than $\sim 10\,\msun$, after hydrogen exhaustion the core contracts very rapidly, out of thermal equilibrium, shell ignition is violent and there is no approach the thermal equilibrium in the envelope which instead is immediately pushed into runaway expansion.
Still, the recombination wave starts from near the surface adding the usual thermal instability to an already complex structure which is out of thermal equilibrium everywhere, from the core, through the shell and then in the envelope.

 So, whether it is the core or the envelope that comes first, cannot be reduced to a chicken and egg problem. During most of the evolution of a star it is the stellar core that leads and the envelope follows,  struggling to adjust. But when envelope thermal instabilities develop, and they do, it is the envelope itself that drives further changes, including those deep in the shell, and does so with the short, thermal timescale. The relative roles of core and envelope in determining the transition to red giants where clearly stated in the conclusions of our 1992 paper, where it was said: ``We have demonstrated that stars become red giants in response to the increasing luminosity being provided by the core, and that the runway expansion -when it takes place- is triggered by the thermal conductivity of the envelope reaching a maximum and then decreasing.  The decrease of thermal conductivity  is caused by the opacity increase promoted by the recombination of heavy ions in the envelope, as the envelope itself expands and cools." (Point 1 of the concluding section). Ultimately, it is the core that drives the envelope to the edge of its catastrophic thermal instability, which however first erupts near the surface and then drills through the star until it reaches to expand even the burning shell itself. All this can be easily verified, suffice to look, and with humility pay attention, at what happens inside models in stellar evolutionary sequences.


\section{Stars that do not become red giants}
\label{sec:not}
The statement above ``... the post-main-sequence developments of all stars  -low- mass, intermediate-mass and high-mass- as they expand to become giants ..."  \citep{faulkner05} is not correct. Not all stars become red giants during their post-MS phase. Section 5.3 of our 1992 paper was dedicated to the effect of metallicity on the evolution to red giant configurations. If metal opacity is the culprit, one expects a big effect of metal abundance on the phenomenon. And indeed the effect is big. Even before 1992, it was known that intermediate- and high-mass stars of low (or zero) metallicity do {\it NOT}  become red giants during their post-MS 
phase, but fail to incur in the thermal instability and ignite helium in the core as blue giants. In this respect, as evidence we quoted \cite{stothers77}, \cite{tornambe84} and \cite{ bertelli85}, see also \cite{cassisi93}. These stars were developing  a large central concentration, a steep density and molecular weight gradient at the edge of their hydrogen-exhausted core, ignited the shell, reached and surpassed the Sch\"onberg-Chandrasekhar limit, just like their more metal rich stars, but failed to become red giants. Perhaps it is worth repeating here that ``We emphasize that in no other proposed explanation of {\it why stars become red giants} one has ever attempted to answer the question `{\it why very metal poor (intermediate-mass) stars do NOT become red giants?'}, a question which instead finds its most natural answer in the frame of our physical interpretation" \citep{renzini92}. And this remains fully valid even to these days, thirty years later, in particular for those articles that have dismissed our demonstration as reported in the previous sections.

\begin{figure}
\includegraphics[width=84mm]{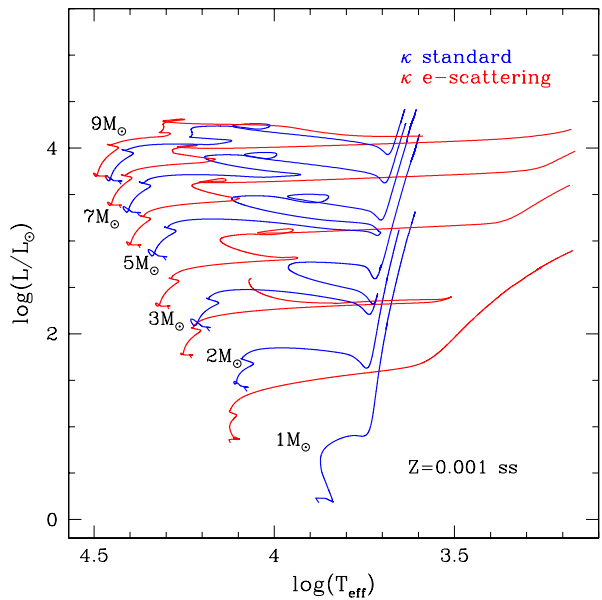}
\caption{Evolutionary sequences for the indicated masses, from the pre-MS all the way to past helium exhaustion at the center and helium shell burning. In blue are the tracks for standard opacities. and metallicity $Z=0.001$. In red are the sequences have assumed that the only source of opacities is pure electron (Thomson) scattering. Courtesy of Santi Cassisi.}
\label{fig:hrd}
\end{figure}

Just to illustrate this further, Figure 1 shows several stellar evolutionary sequences, as described in the caption. Let us consider first the (blue) tracks computed with standard opacities, and relatively low metallicity, $Z=0.001$. In the $9\,\msun$ sequence helium is ignited at the center while the star is still a blue giant, with an effective temperature of $\sim 13,000$ K, when it begins slowly contracting and spends all the core helium-burning phase in the blue. It is not before central helium exhaustion and helium-shell ignition (corresponding to the prominent loop in the track) 
that the fast excursion to the red begins, with the onset of the thermal instability signalled by the luminosity drop past the loop. In practice, it is only after central helium exhaustion that the luminosity grows high enough to trigger the envelope instability, whereas past hydrogen exhaustion the luminosity had not reached such threshold, hence failed to trigger the instability.
This illustrates the point made before that what matters to the envelope is only the {\bf luminosity} at its base, irrespective of the structure inside the outer(most) shell, rather than unspecified 
``developments in the deepest and densest interiors of the stars", as advocated by Faulkner. In other words, there is no {\it subterranean} communication between the core and the envelope.

The case of the $7\,\msun$ track is quite similar, with the main difference being  that helium ignites while the star has reached the slightly cooler temperature of $\sim 9,000$ K. In the case of the $5\,\msun$ star, instead, the thermal instability clearly erupts and the star inflates  to red giant dimension, though an RGB phase is promptly aborted by helium ignition. The $3\,\msun$ sequence is much similar to the case of more metal rich stars, with inflation to the RGB and deflation from it that gives raise to the (first) blue loop. (A hint for the second blue loop is barely visible shortly after the arrival on the RGB.) The $2\,\msun$ star develops the thermal instability and its helium core  becomes electron degenerate, which allows for the extended rise in luminosity terminated by the helium flash.
The extended RGB is even more prominent in the $1\,\msun$ case, where no thermal instability erupts (see next section).

Turning to the red sequences, for which the opacity has been artificially restricted to pure electron scattering, one can notice the following. First, all tracks are systematically brighter and hotter compared to the those with standard opacity, as expected. Past central hydrogen exhaustion, the stellar luminosity steadily increases, no thermal instability sets in and helium is ignited while the star is still at a high effective temperature, that decreases with decreasing stellar mass. Fast excursions to low effective temperatures start only after  helium-shell ignition, corresponding  the (second) blue loops. Moreover,
no envelope convection sets in (opacity is too low) and tracks run to very low effective temperatures. Being now globally  hotter, the $2\,\msun$ star does not develop an electron-degenerate helium core, and helium is ignited under non-degenerate conditions. This pure Thompson-scattering experiment shows that stellar models can be constructed whereby these models reach large dimensions even in absence of a thermal runway instability, suffice for them to become bright enough. Yet, this happens only after helium exhaustion in the core, but not after central hydrogen exhaustion. However, this is not what real stars do, as their opacity is not limited to electron scattering. This experiment shows that (on the computer) a giant configuration can always be achieved, provided that enough luminosity is {\it pumped}  into the envelope. Incidentally, it also shows that a proper RGB is produced only if using realistic opacities.
 

\section{Stars that become red giants without ever breaking their thermal equilibrium}
\label{sec:low}
One objection heard frequently was: {\it but low mass stars do become red giants and yet they do not develop a thermal instability.} True! Suffice to look at the colour-magnitude diagrams of galactic globular clusters, where main sequence, turnoff and subgiant branch join smoothly to the RGB. Most recently, \cite{miller22} states: ``the fact that low mass red giants evolve in a nuclear timescale and develop the most extreme case of giantness, clearly show that thermal instabilities in the envelope are not what pushes stars into red giant dimensions." True, for low-mass stars, but not true for intermediate-mass ones, just more massive than, say $\sim 1.1-1.2\,\msun$, which instead  do {\it become} red giants as a result of the thermal instability.

This point was addressed in our 1992 paper (in Section 5.2) where we said: ``In general, the violence of the runaway decreases with decreasing mass...It does so to the extent that for masses below $\sim 1\,\msun$ (the actual value depends on metallicity) the drop in surface luminosity ... vanishes, and so does the runaway expansion itself. Low mass stars become red giant without ever departing from TE... This follows from the fact that low-mass stars begin their evolution already close to the Hayashi line  and their envelope becomes convective {\it before} the thermal instability has a chance to take place: the early establishment of convection suppresses the thermal instability of the envelope."

So, why do they {\it become} red giants? The reason was already mentioned in Section 3, when saying:  ``In general, when the core luminosity increases the star expands, when it decreases the star contracts." And the core luminosity does indeed steadily increase and does so by about three orders of magnitude (see Figure 1): as the shell around the electron-degenerate helium core burns through, the core mass increases and core radius decreases (this is what happens to degenerate cores and white dwarfs alike). Hence, gravity, pressure, density and temperature in the shell all increase and so does the nuclear energy generated there. To accommodate the increasing luminosity from the core, and radiate it away, the envelope is forced to expand. Contrary to the case of stars experiencing the thermal instability, the core now strongly holds the {\it lead} all the way through --from the MS to the RGB tip-- and the envelope {\it lags} behind. Again, the key factor for the expansion is the luminosity and its steady increase, something that is pathetically absent in the $UV$ homology invariants.

\section{Summary and Conclusions}
\label{sec:summary}
In this paper we have quoted in full the specific criticisms that have been moved to our answer to the question of ``why stars {\it become} red giants?"
It can be appreciated that these dismissals are embarrassingly shallow, being typically limited to just a few, sometimes even cryptic  words.
Several authors admit that a thermal instability develops in the envelope, but just state that it would not be responsible for the runaway expansion. They don't explain why this would not be the case nor what other physical process would drive the expansion instead. Another author admits that stars sweeping to become red giant are indeed in {\it thermal imbalance} but this would be  not caused by a thermal instability of the envelope. Without guessing what other physical process would trigger the thermal imbalance itself.
\cite{faulkner97}, at the beginning of his one page article says: ``Thermal imbalance or even thermal instabilities may in some cases describe {\it how}, but they do not tell us {\it why} stars expand to giant dimensions." As far as physics is concerned, I don't see much difference between the {\it how} and the {\it why}, and Faulkner did not try to explain what the difference would be. In any event, I am satisfied for having proposed how the expansion takes place, and if others believe that a different, whimsical {\it why} exists, and I don't think it does, then I am happy to pass the hand to $UV$-plane explorers. Ultimately, distinguishing evolutionary phases in thermal equilibrium from those that are thermally unstable is critical to properly understand stellar evolution and so the shapes of evolutionary sequences in the HR diagram.

\section*{Acknowledgments} 
I would like to thank Santi Cassisi for his critical reading of the manuscript and for constructive comments. I also thank him for having computed on my request the evolutionary tracks shown in Figure 1.

\section*{Data Availability}
No new data were generated or analyzed in support of this research.

\author[0000-0002-7093-7355]{A.\,Renzini}

\bigskip

\label{lastpage}

\end{document}